\documentclass[final,3p,times,twocolumn]{elsarticle}

\usepackage{amssymb}
\usepackage{bm}
\usepackage{amsmath}
\usepackage{color}
\usepackage[colorlinks,allcolors=blue]{hyperref}
\usepackage{aas_macros}

\newcommand{\bx}{\bf x}

\newcommand{\pfrac}[2]{\left(\frac{#1}{#2}\right)}
\newcommand{\ma}{m_{\rm a}}
\newcommand{\fa}{f_{\rm a}}
\newcommand{\MAS}{M_{\rm AS}}

\newcommand{\fAS}{f_{\rm AS}}
\newcommand{\diff}{{\rm d}}

\journal{Physical Letter B}

\begin{document}

\begin{frontmatter}

\title{Possible evidence of axion stars in HSC and OGLE microlensing events}

\author[ipmu,utokyo]{Sunao Sugiyama\corref{sunaoinfo}}
\ead{sunao.sugiyama@ipmu.jp}
\author[ipmu]{Masahiro Takada}
\author[ipmu,ucla]{Alexander Kusenko}

\affiliation[ipmu]{Kavli Institute for the Physics and Mathematics of the Universe (WPI), UTIAS The University of Tokyo, Kashiwa, Chiba 277-8583, Japan}
\affiliation[utokyo]{Department of Physics, The University of Tokyo, 7-3-1 Hongo, Bunkyo-ku, Tokyo 113-0033 Japan}
\affiliation[ucla]{Department of Physics and Astronomy, University of California, Los Angeles Los Angeles, California, 90095-1547, USA}

\cortext[sunaoinfo]{Corresponding author}

\begin{abstract}
    Dark matter in the form of axions is expected to form axion stars.  Such axion stars could be discovered by microlensing events.  In particular, some candidate events reported by Subaru HSC and OGLE can be explained simultaneously if  the axion  stars  with masses of the order of the Earth mass  make  up  about $\sim27^{+7}_{-13}$
    percent
    of dark matter.  For QCD axions, this corresponds to the axion mass in the range $10^{-9}-10^{-6}$ eV, which is consistent with the experimental constraints, as well as the cosmological anthropic window of parameters. 
\end{abstract}


\end{frontmatter}

\section{Introduction}
A natural solution to the strong CP problem is associated with the existence of a light scalar field, the axion~\cite{Peccei:1977hh,Weinberg:1977ma,Wilczek:1977pj}.  The axion particles can be produced in the early universe, and they can account for all or part of dark matter~\cite{Preskill:1982cy, Abbott:1982af, Dine:1982ah}.  In addition, string theory predicts a number of axion-like particles, which are not necessarily associated with the strong  interactions~\cite{Svrcek:2006yi,Arvanitaki:2009fg,Conlon.Conlon.2006}. Dark matter in the form of axions is expected to form axion stars~\cite{Tkachev:1991ka,Kolb:1993zz,Kolb:1995bu,Guth:2014hsa,Schiappacasse:2017ham,Levkov:2018kau, Visinelli:2017ooc,Seidel.Suen.1993, Schive.Broadhurst.2014,Eggemeier.Niemeyer.2019,Chen.Niemeyer.2020}. Microlensing observations can be used to set upper limits on the abundance of such compact objects~\cite{Croon:2020ouk, Croon:2020wpr,Fujikura.Yamaguchi.2021}.

We study whether microlensing events with  short timescale light curves, recently reported in Refs.~\cite{Niikura:2017zjd,Niikura:2019kqi}, can be explained by axion stars, and we infer the corresponding axion parameters.  

\section{Stable solution of axion star}
Axions, or  more generally axion-like particles, are described in field theory by a real scalar field $\phi(t,\bx)$  with the following potential: 
\begin{align}
V(\phi)&=
\Lambda^4\left[1-\sqrt{1-\frac{4m_{\rm u}m_{\rm d}}{(m_{\rm u}+m_{\rm d})^2}\sin^2\left(\frac{\phi}{2f_{\rm a}}\right)}\right]\nonumber\\
&\simeq \frac{m_a^2}{2}\phi^2-\frac{m_a^2}{24f_a^2}\gamma\phi^4+...,
\label{eq:V_axion}
\end{align}
where $m_{\rm u}\simeq2.2~{\rm MeV}$, $m_{\rm d}\simeq4.7~{\rm MeV}$, and $f_a$ are the up and down quark masses and the axion decay constant, respectively. 
The overall scale of the potential is related to the pion mass and the pion decay constant \citep{GrillidiCortona:2015jxo}: $\Lambda^4=f_{\pi}^2m_{\pi}^2$ where $f_{\pi}\simeq92~{\rm MeV}$ and $m_\pi\simeq135~{\rm MeV}$.
In the second line, we approximated the potential by a series expansion around $\phi=0$ assuming $|\phi|\ll f_a$. 
The first term in the expanded potential acts as an effective mass term $m_a=\sqrt{m_{\rm u}m_{\rm d}/(m_{\rm u}+m_{\rm d})^2}\Lambda^2/f_a$. 
The second term represents an attractive scalar self-interaction, which, together with self-gravity, can lead to an instability and formation of axion stars~\cite{Tkachev:1991ka,Kolb:1993zz,Kolb:1995bu,Guth:2014hsa,Schiappacasse:2017ham,Levkov:2018kau, Visinelli:2017ooc}. 
Due to the presence of the factor $\gamma\equiv1-3m_{\rm u}m_{\rm d}/(m_{\rm u}+m_{\rm d})^2$ in the second term, for which we will use the fiducial value $\gamma\sim 0.34$ in this letter, the field has the effective decay constant as $f_{\rm a}'=f_{\rm a}\gamma^{-1/2}$.
The fraction of cosmologically produced axions that end up in axion stars is subject of ongoing investigation. The axion stars can be stable, or, if they reach some critical mass, they can collapse emitting a burst of relativistic axions~\cite{Levkov:2016rkk,Eby:2021ece,Helfer.Becerril.2017,Michel.Moss.2018}. Among the stable solutions of axion stars, the most massive and compact one has the mass and radius given as ~\cite{Schiappacasse:2017ham, Visinelli:2017ooc}
\begin{align}
    M_{\rm AS} &= 1.2\times10^{-6}M_\odot\pfrac{\ma}{10^{-8}{\rm eV}}^{-1}\pfrac{\fa}{10^{14}{\rm GeV}}\label{eq:axion-star-mass},
\end{align}
and
\begin{align}
    R_{\rm AS} &= 7.8\times10^2 {\rm km} \pfrac{\ma}{10^{-8}{\rm eV}}^{-1}\pfrac{\fa}{10^{14}{\rm GeV}}^{-1}\label{eq:axion-radius}.
\end{align}
Here we adopted, as an example, the axion parameters  $m_a=10^{-8}~{\rm eV}$ and $f_a=10^{14}~{\rm GeV}$ that are consistent with the QCD axion.\footnote{ We note that the corresponding axion stars have masses close to the Earth mass ($\simeq 3\times 10^{-6}M_\odot$) and a smaller radius than that of the Earth ($\simeq 6400~{\rm km}$).}  The best-motivated mass range for the axions extends above and below this mass.  
If the Peccei-Quinn (PQ) symmetry is broken before inflation, and the axions are produced via the misalignment mechanism, the initial misalignment  $\phi_i$  determines the axion abundance.  If $\phi_i/f_a \sim 1$, there is a limit $m_a > 10^{-5}{\rm eV}$ because the abundance of axion dark matter exceeds the observed values for smaller masses.  However, different parts of the universe (or ``multiverse''), on the scales much greater than today's horizon, could have different values of $\phi_i$~\cite{Linde:1982uu,Wilczek:2004cr}.  Regions with a small misalignment may occupy a small volume of the universe, but they may contain a large fraction of potential observers~\cite{Linde:1982uu,Wilczek:2004cr,Tegmark:2005dy}. This is particularly intriguing in view of the arguments that stars and planets hosting intelligent life could not have formed in a universe with a dark matter content greater than the observed value~\cite{Tegmark:2005dy}.~\footnote{The increased amount of dark matter causes a change in the matter-radiation equality temperature. 
If the matter comes to dominate the universe too early, the density perturbations grow and become nonlinear before recombination; the baryons and radiation get trapped inside the collapsing halos, and the baryon coupling to photons maintains the Jeans mass at a constant value as the collapse proceeds. 
As a result,  large amounts of coupled baryon-radiation fluid are dragged into the potential wells created by clumps of dark matter, leading to a universe with supermassive black holes, photons, and neutrinos, but without stars and planets~\cite{Tegmark:2005dy}. 
The same arguments apply to some other kinds of dark matter, such as, e.g., moduli~\cite{Kusenko:2012ch}.}  
Alternatively, the small axion mass can be reconciled with the dark matter abundance if some additional particle, a {\em diluton} decays and produces entropy, diluting the axion density, see, e.g., Refs.~\cite{Fuller:2011qy,Patwardhan:2015kga,Hasegawa:2019jsa}. 

In the scenario of the PQ symmetry breaking before inflation, there is a potential problem with isocurvature perturbations in excess of the existing observational bounds~\cite{AXENIDES1983178,LINDE1985375,LYTH1990408,LINDE1990353,Turner:1990uz,PhysRevD.32.3178,1991PhLB..259...38L}. This problem can be solved by a relatively low scale of  inflation \citep{Visinelli:2009kt,Kawasaki:1995vt,Kobayashi:2013nva,Takahashi:2018tdu}. In addition, if the reheating temperature is sufficiently low and there is significant entropy production, the axion dark matter is consistent with $f_a \lesssim 10^{13} {\rm GeV}$~\citep{Visinelli:2009kt,Kawasaki:1995vt,Schiappacasse:2021zlr}, and the upper bound can be even higher, up to $10^{14}\, {\rm GeV}$, because of the uncertainties in the modeling of axion strings, their evolution, and the spectrum of emitted axions. 
Thus we do not impose the cosmological bound of $10^{-5}\,{\rm eV}$ on the axion mass.

\section{Gravitational microlensing}
When an axion star and a background star are aligned along the line-of-sight direction of an observer,
the star is multiply imaged by strong lensing \citep{1986ApJ...304....1P}. The multiple images are usually not resolved by a telescope, and an observer can identify this lensing event from a  time-varying brightness of the same star, 
which forms a characteristic light curve, because the source star, lens (here an axion star), and an observer have the relative motion.
This is the so-called microlensing. The length scale characterizing the cross section of microlensing is the Einstein radius, given by 
\begin{align}
R_{\rm E}&=\frac{\sqrt{4GM D}}
{c} \nonumber\\
 &\simeq 1.6\times 10^6~{\rm km}\left(\frac{M}{1.2\times 10^{-6}M_\odot}\right)^{1/2}\left(\frac{D}{4~{\rm kpc}}\right)^{1/2},
 \label{eq:RE}
\end{align}
where $D\equiv d_{\rm l}d_{\rm ls}/d_{\rm s}$ ($d_{\rm l}, d_{\rm s}$ and $d_{\rm ls}$ are distances to lens, to source and between lens and source from an observer, respectively). If separation between a background star and a lens is smaller than $R_{\rm E}$ on the sky, the microlensing occurs. 
Comparing Eqs.~(\ref{eq:axion-radius}) and (\ref{eq:RE}) manifests $R_{\rm E}\gg R_{\rm AS}$ for an axion star with $M_{\rm AS}=1.2\times 10^{-6}M_\odot$ corresponding to the axion parameters taken in the equations, meaning that such an axion star is sufficiently compact to cause a microlensing event.

In this paper we assume that axion stars are formed in the early universe and constitute some mass fraction of dark matter (DM) that exists in the Milky Way (MW) (and also the Andromeda galaxy). A shape of the mass function of axion stars, even if formed, depends on details of physics inherent in the formation process of axion stars in the early universe. Generally speaking, axion stars with masses smaller than the mass scale (Eq.~\ref{eq:axion-star-mass}) can be formed, and  the mass function can be extended to smaller mass scales. In this paper, for simplicity, we assume a monotonic (delta-function) mass function of a single mass scale $M_{\rm AS}$ for axion stars, and that the mass fraction of axion stars to total DM in the MW is parameterized by $f_{\rm AS}$. The result with a monochromatic mass function in this paper can be always re-interpreted into general mass functions by convolving our result with a mass function of interest \cite{Inomata:2017okj}.

\section{Microlensing constraints on axion parameters}
To constrain parameters of axion stars, we use the published results of microlensing observations 
in the Subaru HSC data \citep{Niikura:2017zjd} and in the Optical Gravitational Lensing Experiment (OGLE) \cite{2017Natur.548..183M}. 

\citet{Niikura:2017zjd} used the dense-cadence (2~min sampling rate), 7-hour long observation data of the Andromeda galaxy (M31) with Subaru HSC, and reported one possible microlensing event with short timescale of $\sim 1$~hour, implying a lens with moon mass scale ($\sim 10^{-8}M_\odot$) \citep[also see][]{Sugiyama.Takada.2020}.
The direction of M31 is through the halo region of the MW, and we assume that the HSC microlensing event is due to an axion star that has the same spatial and velocity distributions in the MW and M31 halo regions as those implied from the DM halo model \citep[see Ref.][for the details]{Niikura:2017zjd}.

Ref.~\citep{2017Natur.548..183M} used the publicly-available 2622 microlensing events, obtained from the 5-year OGLE observation of $\sim 5\times 10^7$ stars in the Galactic bulge region, to discuss the populations of lensing objects. The OGLE events revealed a distinct population of 6 ultra-short timescale microlensing events ($[0.1,0.3]$ day scales) compared to the other majority of events that have a smooth distribution of microlensing timescales and are explained by standard populations of lensing objects such as brown dwarfs, main-sequence stars and white dwarfs in the standard model of the MW \citep[also see][]{2003ApJ...591..204S,2011Natur.473..349S,2017Natur.548..183M,Abrams:2020jvs,2021arXiv210313015T}. 
\citet{Niikura:2019kqi} discussed that these ultra-short timescale events can be explained by compact objects such as primordial black holes that have the same spatial and velocity distributions as predicted in the MW halo model, if the compact objects are in the range of Earth mass scale ($\sim 10^{-6}M_\odot$). Since the OGLE observation is towards the Galactic bulge, the compact objects might be due to free-floating planets that are in the Galactic disk and bulge regions. Even in this case, theory needs to explain why free-floating planets have a preferred mass scale, i.e. Earth mass scale, rather than a smooth mass distribution. 
In this paper we use 1, 2, 2 and 1 microlensing events in each of 4 bins of the timescales $t_{\rm E}=[0.1,0.3]~{\rm days}$ as given in Fig.~5 of \citet{Niikura:2019kqi}, assuming that the events are due to axion stars. Here axion stars are distributed within the entire MW halo region extending to 
$\sim 200~{\rm kpc}$ in radius from the Galactic center, and the microlensing occurs if an axion star comes across 
the Galactic disk and bulge regions and passes through in front of a background star in the Galactic bulge region along the line-of-sight direction of an observer.

\begin{figure}[tb]
\centering
  \centering
  \includegraphics[width=0.45\textwidth]{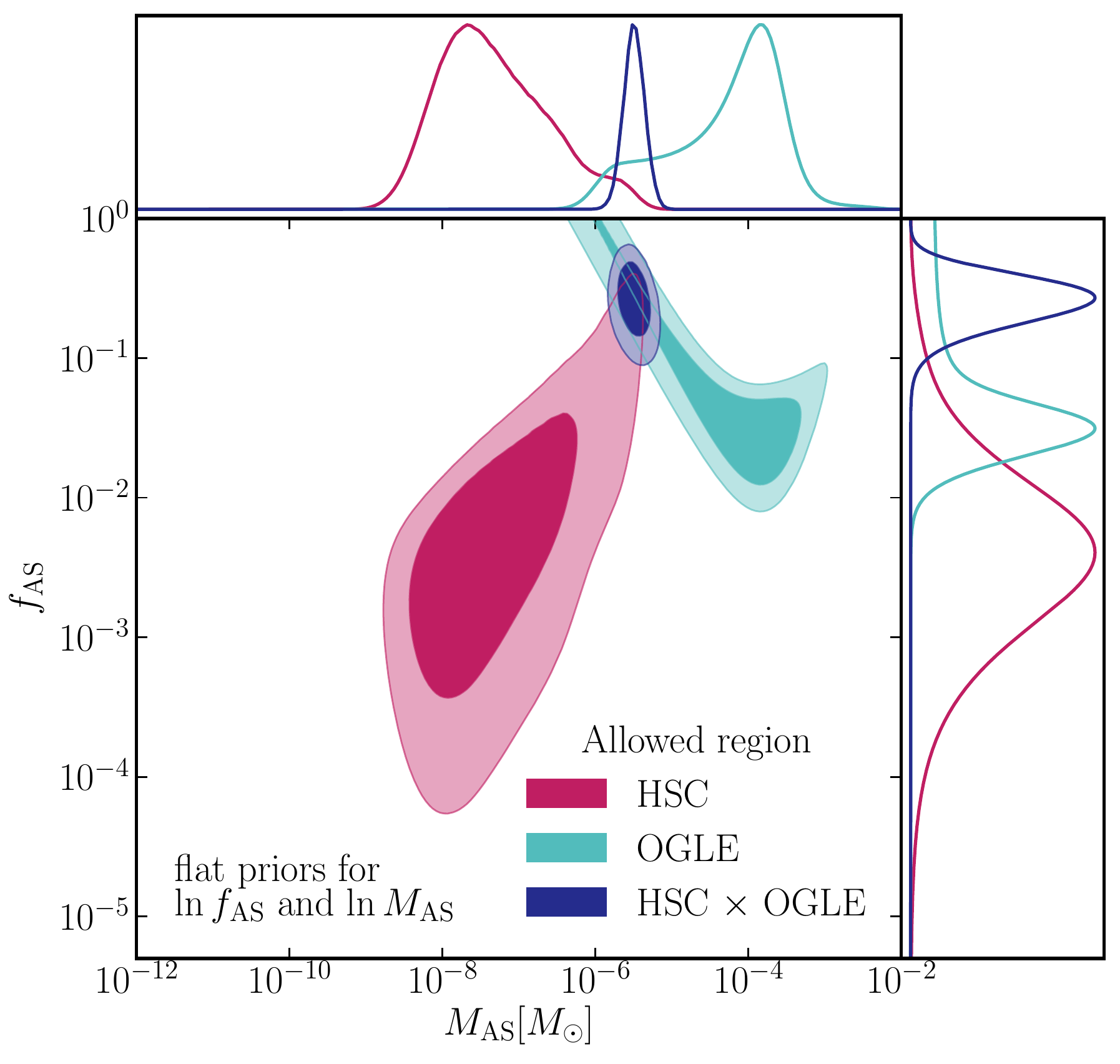}
  \caption{Posterior distribution of axion star mass ($\MAS$) and the mass fraction ($\fAS$) of axion stars to the dark matter of the MW and M31 halo regions, which are obtained assuming that the one possible microlensing event of the HSC M31 observation \cite{Niikura:2017zjd} and the six ultra-short timescale microlensing events of OGLE \cite{2017Natur.548..183M,Niikura:2019kqi} are due to axion stars. 
  Red and cyan regions are 68\%(95\%) confidence regions that are consistent with the combined HSC and OGLE events, respectively. 
  Blue region is the {\it joint} posterior allowed region for the HSC and OGLE events. 
  Upper (right) panel shows the 1D posterior distributions of $M_{\rm AS}$ ($f_{\rm AS}$).
  }
  \label{fig:allowed-MAS-fAS}
\end{figure}
Assuming that axions stars follow the spatial and velocity distributions predicted by the standard DM halo model for the MW and M31, we can compute the expected number of microlensing events in a given bin of the timescale bin $t_{\rm E}=[t_{{\rm E},i},t_{{\rm E},i+1}]$, for a given microlensing observation (either of the Subaru HSC or OGLE observation) 
\citep[see][for details of the equations used in the computation]{Niikura:2017zjd,Niikura:2019kqi}:
\begin{align}
    N_{\rm exp}(t_{{\rm E},i}) = t_{\rm obs}N_{\rm s}\fAS\int_{t_{{\rm E},i}}^{t_{{\rm E},i+1}}\diff t_{\rm E}
    \left.\frac{\diff \Gamma}{\diff t_{\rm E}}\epsilon(t_{\rm E})\right|_{M_{\rm AS}}\ , 
    \label{eq:Nexp}
\end{align}
where $t_{\rm obs}$ is a duration of the microlensing observation, and $N_{\rm s}$ is the number of monitored stars used in the microlensing observation. $\diff\Gamma/\diff t_{\rm E}$ is the {\it differential} event rate giving the expected number of microlensing events of a given timescale $t_{\rm E}$ per unit observation time and per a single source star; the dimension is [${\rm events}~{\rm sec}^{-2}$]. 
$\epsilon(t_{\rm E})$ is the detection efficiency of microlensing that quantifies the probability that a microlensing event of timescale $t_{\rm E}$ is successfully recovered (detected) by the observation, which was estimated in Refs.~\cite{Niikura:2017zjd} and \cite{2017Natur.548..183M} for the Subaru HSC and OGLE data, respectively. For the halo models of the MW and M31, we assume a Navarro-Frenk-White model \cite{1997ApJ...490..493N} given in Ref.~\cite{Klypinetal:02} that reproduces the observations such as the rotation curve. For axion stars we employ the two parameters, $M_{\rm AS}$ and $\fAS$: $M_{\rm AS}$ is the mass of axion star, and $\fAS$ is a parameter to model the mass fraction of axion stars to the total DM mass of the MW halo (and the M31 halo). Note $\mathrm{d}\Gamma/\mathrm{d}t_{\rm E}\propto 1/\MAS$ because the number density distribution of axion stars in the MW (and M31) halo region is given by $n_{\rm AS}({\bf x})\propto \rho_{\rm NFW}({\bf x})/\MAS$, where $\rho_{\rm NFW}({\bf x})$ is the DM radial profile with respect to the Galactic or M31 center, as predicted by the NFW model of the MW and M31. 
When we evaluate the expected number of events, we take into account the finite source size effect of microlensing in an optical wavelength observation, which reduces the expectation number of events for axion stars on low mass scales $M_{\rm AS}\lesssim 10^{-7}M_\odot$ \citep{Sugiyama.Takada.2020}.

\begin{figure*}[tb]
  \centering
  \includegraphics[width=0.9\textwidth]{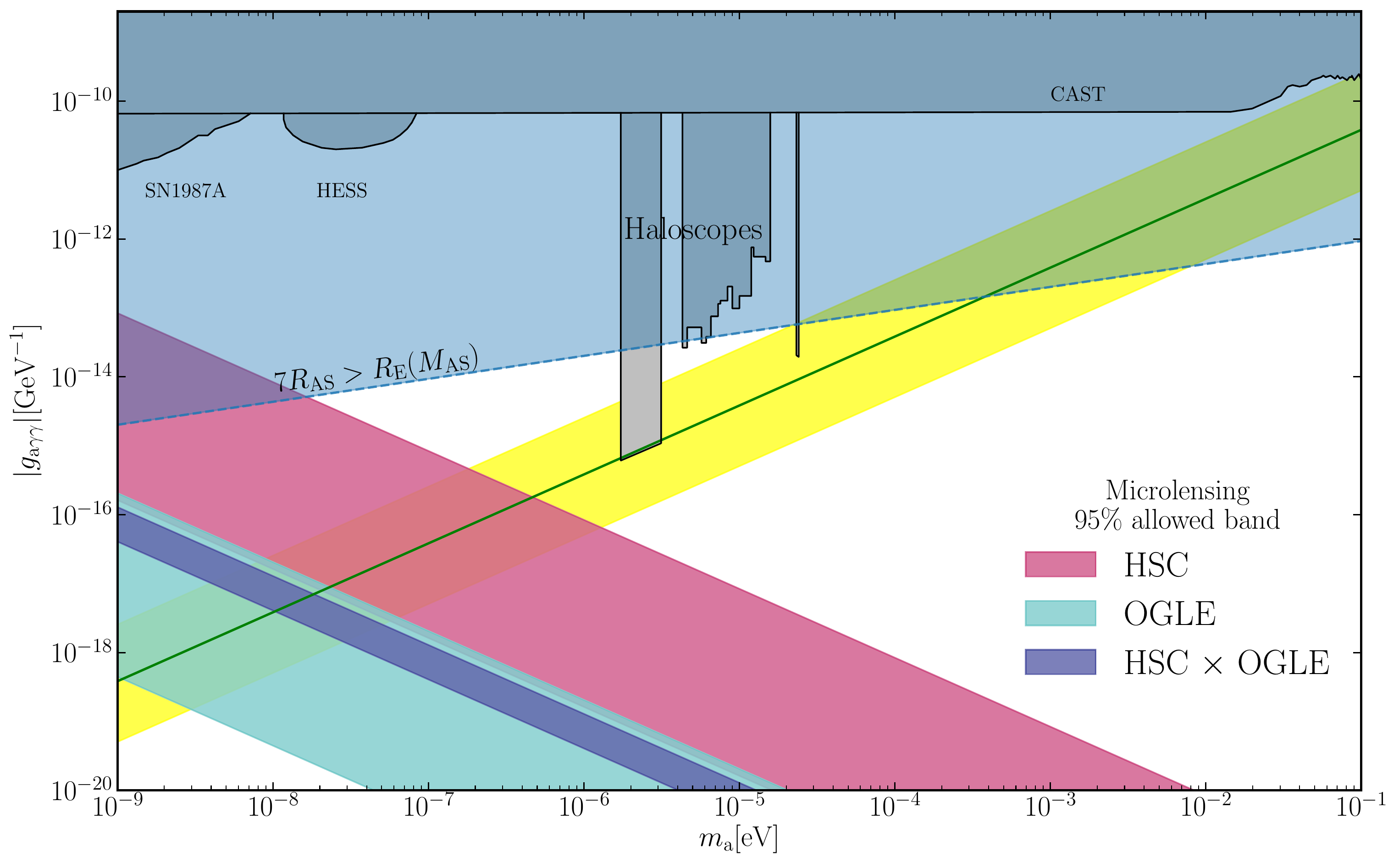}
  \caption{Allowed regions in the axion parameter space, the axion mass scale ($m_{\rm a}$) and the axion-photon coupling constant ($|g_{\rm a\gamma\gamma}|$), obtained from the allowed regions of axion star mass in Fig.~\ref{fig:allowed-MAS-fAS} using Eq.~(\ref{eq:axion-star-mass}). 
  Color scheme is similar to that of Fig.~\ref{fig:allowed-MAS-fAS}. Blue color bands are the allowed regions for the joint constraints of the HSC and OGLE results. Green line and yellow band denote a range of the QCD axion parameters \citep[e.g.][]{DiLuzio:2016sbl}, displaying an overlapping region with the microlensing allowed regions around $m_{\rm a}\sim 10^{-8}~{\rm eV}$. Note that axion stars with masses lighter than the scale indicated by the microlensing constraints can be stable and exist (see text for details). The shaded region, labeled as ``$7R_{\rm AS}>R_{E}(M_{\rm AS})$'', denote the range of axion parameters where the axion star is {\it not} compact enough to be regarded as a point lens object \citep{Fujikura.Yamaguchi.2021}.
  Some bounds may be altered if a large fraction of the axion dark matter resides in axion stars. 
  For comparison the existing upper limits on the coupling constant by the terrestrial experiments and previous works are shown in gray: SN1987A \cite{2008LNP...741...51R}, HESS \cite{HESS:2013udx}, Haloscopes \cite{2014IJMPA..2943004S}, and CAST \cite{2007NJPh....9..169K, CAST:2017uph}.
  We assume all the galactic dark matter is in the form of diffuse axion in the Haloscope limits rather than shifting them by our estimation $1-f_{\rm AS}$.
  }
  \label{fig:allowed-ma-fa}
\end{figure*}

We perform parameter estimation based on the Bayesian inference for parameter set $\bm{p}=\{\MAS,\fAS\}$:
\begin{align}
    P(\bm{p}) \propto \mathcal{L}(\bm{d}|\bm{p})
    \Pi(\bm{p}),\label{eq:posterior}
\end{align}
where $\mathcal{L}$ is the likelihood and $\Pi$ is the prior. For the likelihood, we can safely assume that the observed number of microlensing events in a given timescale bin follows the Poisson distribution, because different microlensing events are safely considered independent due to the smallness of the microlensing optical depth for each of source stars:
\begin{align}
    {\cal L} = \prod_{i} \frac{(N_{{\rm exp},i})^{N_i}}{(N_i)!}{\rm e}^{-N_{{\rm exp},i}} \ , \label{eq:poisson-likelihood}
\end{align}
where $N_i$ is the number of observed microlensing events in the $i$-th timescale bin ($t_{{\rm E},i}$), and $N_{{\rm exp},i}$ is the expectation value predicted from the model assuming the abundance of axion stars in the MW and M31 halo regions (Eq.~\ref{eq:Nexp}). As described above, we consider one bin and four bins of the microlensing light curve timescales for the HSC and OGLE events, respectively. 
We employ flat priors in the logarithmic scale for $\fAS$ and $\MAS$.

\begin{table}[t]
    \centering
    \caption{Marginalized constraint of axion star parameters. The mode values are presented with 68\%  highest density interval for each dataset.}
    \renewcommand*{\arraystretch}{1.4}
    \begin{tabular}{c|cc}\hline\hline
                         & $\log [M_{\rm AS}/M_\odot]$       & $\log f_{\rm AS}$\\
        \hline
        HSC              & $-7.68_{-0.50}^{+0.85}$ & $-2.39_{-0.66}^{+0.61}$ \\
        OGLE             & $-3.86_{-0.85}^{+0.45}$ & $-1.50_{-0.32}^{+0.37}$ \\
        HSC$\times$OGLE  & $-5.52_{-0.10}^{+0.15}$ & $-0.57_{-0.18}^{+0.16}$ \\
        \hline\hline
    \end{tabular}
    \renewcommand*{\arraystretch}{1}
    \label{tab:marginalHDI}
\end{table}
The contours in Fig.~\ref{fig:allowed-MAS-fAS} shows the 68\% and 95\% credible regions on $M_{\rm AS}$ and $f_{\rm AS}$, obtained from either of the HSC or OGLE short timescale microlensing events or the joint constraints, respectively. 
The marginalized constraints on $f_{\rm AS}$ and $M_{\rm AS}$ are summarized in Table~\ref{tab:marginalHDI} for each posterior.
To be more precise, $\log f_{\rm AS}=-0.57_{-0.18}^{+0.16}$ and $\log [M_{\rm AS}/M_\odot]=-5.52_{-0.10}^{+0.15}$ for the HSC and OGLE events, respectively.
It is intriguing that the two independent microlensing events of HSC and OGLE are consistent with each other, if the axion stars have an Earth mass scale ($10^{-6}M_\odot$) and make up about a few tens percent for the mass fraction to the total DM. 
(A similar conclusion can be applied to primordial black holes \cite{Niikura:2019kqi}.)

We also note that a slight difference between the HSC and OGLE parameter ranges in Fig.~\ref{fig:allowed-MAS-fAS} does not necessarily imply a ``tension'' of these observations. The two sets of observations are sensitive to different mass ranges due to the difference in the observational time scales. The single event from HSC favors the mass scale $M_{\rm AS}\sim[10^{-9},10^{-6}]M_\odot$, but this detection has no implications for the OGLE sensitivity range, $M_{\rm AS}\sim[10^{-6}, 10^{-3}]M_{\odot}$ where the other six events are detected. If one assumes a monochromatic mass function, as we have done for simplicity, then the overlap region is favored by the combination of HSC and OGLE data. However, many models predict an extended mass function, as e.g.  Ref.~\cite{2020PhRvL.125r1304K}. A broader mass function can simultaneously be consistent with the HSC and OGLE events as the two observations probe different parts of the mass function.

We then discuss implications of the results in Fig.~\ref{fig:allowed-MAS-fAS} on axion parameters. By marginalizing the posterior distribution $P(M_{\rm AS},f_{\rm AS})$ over $f_{\rm AS}$, we can obtain the projected posterior distribution of $M_{\rm AS}$. 
We then convert the credible interval of $M_{\rm AS}$ to the posterior region of axion parameters using 
Eq.~(\ref{eq:axion-star-mass}). Here we consider, for the axion parameters, $m_a$ and $g_{\rm a\gamma\gamma}$, where $g_{\rm a\gamma\gamma}$ is the axion and photon coupling constant. We used the relation  $g_{\rm a\gamma\gamma}=\alpha_{\rm EM}/(2\pi\fa)(E/N-1.92)$, where $\alpha_{\rm EM}=1/137$ is the fine-structure constant of electromagnetic interaction, and $E$ and $N$ are anomaly coefficients depending on the QCD model.
We use KSVZ model, where $E/N=0$, as a benchmark model. We note that an uncertainty in $E/N$ alters $g_{\rm a\gamma\gamma}$ by up to a factor of 10 \cite{DiLuzio:2016sbl}.

Fig.~\ref{fig:allowed-ma-fa} shows that the Subaru and OGLE ultra-short-timescale microlensing events are consistent with an  interesting range of axion parameters, including 
the QCD axions in the mass range $m_{\rm a}\sim 10^{-8}~{\rm eV}$. It is also interesting to note that the degeneracy direction of the microlensing constraint is quite complementary to existing constraints from terrestrial experiments. The region, denoted by ``$7R_{\rm AS}>R_{\rm E}(M_{\rm AS})$'', indicates that the axion star is {\it not} compact enough to be regarded as a point lens object \citep{Fujikura.Yamaguchi.2021}.
We also note that the axion mass range obtained in this paper is complementary to the direct detection of axion-like particle stars probed by optical magnetometer, e.g. the Global Network of Optical Magnetometers to search for Exotic physics (GNOME) collaboration, which has sensitivity around $m_a\gtrsim10^{-4}~{\rm eV}$ \cite{Kimball.Wickenbrock.2018, Pustelny.Budker.2013}.
In this letter, we focus on the most massive and compact axion star solution in Eqs.~(\ref{eq:axion-star-mass}) and (\ref{eq:axion-radius}) among possible stable solutions. Depending on the detail of the axion star formation scenario, lighter and more dispersed axion stars can be formed. In this case, our constraints in $(m_{\rm a}, g_{\rm a\gamma\gamma})$ are shifted toward the lighter axion mass.

\section{Conclusion}
Microlensing events in the HSC and OGLE data are consistent with $\sim27^{+7}_{-13}$ percent of dark matter in the form of axion stars.  If the QCD axions make up all or most of dark matter, it is likely that $\sim27^{+7}_{-13}$ percent of the axions have condensed into axion stars~\cite{Tkachev:1991ka,Kolb:1993zz,Kolb:1995bu,Guth:2014hsa,Schiappacasse:2017ham,Levkov:2018kau}.  For the QCD axion, the detected microlensing events correspond to the masses in the range $10^{-9}-10^{-6}\, {\rm eV} $, which is consistent with the experimental bounds~\cite{Graham:2015ouw}. 
Further microlensing observations such as those using the Subaru HSC and the upcoming Vera C. Rubin Observatory telescope will enable us to test the possible contribution of axion stars to dark matter.

\section{Acknowledgments}
We thank Mark Hertzberg and Kaz Kohri for useful discussions. 
The work of A.K. was supported by the U.S. Department of Energy (DOE) Grant No. DE-SC0009937. A.K. thanks the Aspen Center for Physics, which is supported by National Science Foundation grant PHY-1607611. This work was supported by the World Premier International Research Center Initiative (WPI), MEXT, Japan. The work of S.S. was supported by International Graduate Program for Excellence in Earth-Space Science (IGPEES), World-leading Innovative Graduate Study (WINGS) Program, the University of Tokyo. This research was also supported in part by the National Science Foundation under Grant No. NSF PHY-1748958 and by JSPS KAKENHI Grant Numbers JP15H05893, JP19H00677, JP20H05600, JP20H05853, JP20H05855, and JP21J10314.

\biboptions{sort&compress}
\bibliographystyle{elsarticle-num-names}
\bibliography{bibliography}

\newcommand{\noop}[1]{}
\begin{thebibliography}{69}
\expandafter\ifx\csname natexlab\endcsname\relax\def\natexlab#1{#1}\fi
\providecommand{\url}[1]{\texttt{#1}}
\providecommand{\href}[2]{#2}
\providecommand{\path}[1]{#1}
\providecommand{\DOIprefix}{doi:}
\providecommand{\ArXivprefix}{arXiv:}
\providecommand{\URLprefix}{URL: }
\providecommand{\Pubmedprefix}{pmid:}
\providecommand{\doi}[1]{\href{http://dx.doi.org/#1}{\path{#1}}}
\providecommand{\Pubmed}[1]{\href{pmid:#1}{\path{#1}}}
\providecommand{\bibinfo}[2]{#2}
\ifx\xfnm\relax \def\xfnm[#1]{\unskip,\space#1}\fi
\bibitem[{Peccei and Quinn(1977)}]{Peccei:1977hh}
\bibinfo{author}{R.~D. Peccei}, \bibinfo{author}{H.~R. Quinn},
\newblock \bibinfo{title}{{CP Conservation in the Presence of Instantons}},
\newblock \bibinfo{journal}{Phys. Rev. Lett.} \bibinfo{volume}{38}
  (\bibinfo{year}{1977}) \bibinfo{pages}{1440--1443}.
  \DOIprefix\doi{10.1103/PhysRevLett.38.1440}.
\bibitem[{Weinberg(1978)}]{Weinberg:1977ma}
\bibinfo{author}{S.~Weinberg},
\newblock \bibinfo{title}{{A New Light Boson?}},
\newblock \bibinfo{journal}{Phys. Rev. Lett.} \bibinfo{volume}{40}
  (\bibinfo{year}{1978}) \bibinfo{pages}{223--226}.
  \DOIprefix\doi{10.1103/PhysRevLett.40.223}.
\bibitem[{Wilczek(1978)}]{Wilczek:1977pj}
\bibinfo{author}{F.~Wilczek},
\newblock \bibinfo{title}{{Problem of Strong $P$ and $T$ Invariance in the
  Presence of Instantons}},
\newblock \bibinfo{journal}{Phys. Rev. Lett.} \bibinfo{volume}{40}
  (\bibinfo{year}{1978}) \bibinfo{pages}{279--282}.
  \DOIprefix\doi{10.1103/PhysRevLett.40.279}.
\bibitem[{Preskill et~al.(1983)Preskill, Wise, and Wilczek}]{Preskill:1982cy}
\bibinfo{author}{J.~Preskill}, \bibinfo{author}{M.~B. Wise},
  \bibinfo{author}{F.~Wilczek},
\newblock \bibinfo{title}{{Cosmology of the Invisible Axion}},
\newblock \bibinfo{journal}{Phys. Lett. B} \bibinfo{volume}{120}
  (\bibinfo{year}{1983}) \bibinfo{pages}{127--132}.
  \DOIprefix\doi{10.1016/0370-2693(83)90637-8}.
\bibitem[{Abbott and Sikivie(1983)}]{Abbott:1982af}
\bibinfo{author}{L.~F. Abbott}, \bibinfo{author}{P.~Sikivie},
\newblock \bibinfo{title}{{A Cosmological Bound on the Invisible Axion}},
\newblock \bibinfo{journal}{Phys. Lett. B} \bibinfo{volume}{120}
  (\bibinfo{year}{1983}) \bibinfo{pages}{133--136}.
  \DOIprefix\doi{10.1016/0370-2693(83)90638-X}.
\bibitem[{Dine and Fischler(1983)}]{Dine:1982ah}
\bibinfo{author}{M.~Dine}, \bibinfo{author}{W.~Fischler},
\newblock \bibinfo{title}{{The Not So Harmless Axion}},
\newblock \bibinfo{journal}{Phys. Lett. B} \bibinfo{volume}{120}
  (\bibinfo{year}{1983}) \bibinfo{pages}{137--141}.
  \DOIprefix\doi{10.1016/0370-2693(83)90639-1}.
\bibitem[{Svrcek and Witten(2006)}]{Svrcek:2006yi}
\bibinfo{author}{P.~Svrcek}, \bibinfo{author}{E.~Witten},
\newblock \bibinfo{title}{{Axions In String Theory}},
\newblock \bibinfo{journal}{JHEP} \bibinfo{volume}{06} (\bibinfo{year}{2006})
  \bibinfo{pages}{051}. \DOIprefix\doi{10.1088/1126-6708/2006/06/051}.
  \href{http://arxiv.org/abs/hep-th/0605206}{{\tt arXiv:hep-th/0605206}}.
\bibitem[{Arvanitaki et~al.(2010)Arvanitaki, Dimopoulos, Dubovsky, Kaloper, and
  March-Russell}]{Arvanitaki:2009fg}
\bibinfo{author}{A.~Arvanitaki}, \bibinfo{author}{S.~Dimopoulos},
  \bibinfo{author}{S.~Dubovsky}, \bibinfo{author}{N.~Kaloper},
  \bibinfo{author}{J.~March-Russell},
\newblock \bibinfo{title}{{String Axiverse}},
\newblock \bibinfo{journal}{Phys. Rev. D} \bibinfo{volume}{81}
  (\bibinfo{year}{2010}) \bibinfo{pages}{123530}.
  \DOIprefix\doi{10.1103/PhysRevD.81.123530}.
  \href{http://arxiv.org/abs/0905.4720}{{\tt arXiv:0905.4720}}.
\bibitem[{Conlon(2006)}]{Conlon.Conlon.2006}
\bibinfo{author}{J.~P. Conlon},
\newblock \bibinfo{title}{{The QCD axion and moduli stabilisation}},
\newblock \bibinfo{journal}{Journal of High Energy Physics}
  \bibinfo{volume}{2006} (\bibinfo{year}{2006}) \bibinfo{pages}{078}.
  \DOIprefix\doi{10.1088/1126-6708/2006/05/078}.
  \href{http://arxiv.org/abs/hep-th/0602233}{{\tt arXiv:hep-th/0602233}}.
\bibitem[{Tkachev(1991)}]{Tkachev:1991ka}
\bibinfo{author}{I.~I. Tkachev},
\newblock \bibinfo{title}{{On the possibility of Bose star formation}},
\newblock \bibinfo{journal}{Phys. Lett. B} \bibinfo{volume}{261}
  (\bibinfo{year}{1991}) \bibinfo{pages}{289--293}.
  \DOIprefix\doi{10.1016/0370-2693(91)90330-S}.
\bibitem[{Kolb and Tkachev(1993)}]{Kolb:1993zz}
\bibinfo{author}{E.~W. Kolb}, \bibinfo{author}{I.~I. Tkachev},
\newblock \bibinfo{title}{{Axion miniclusters and Bose stars}},
\newblock \bibinfo{journal}{Phys. Rev. Lett.} \bibinfo{volume}{71}
  (\bibinfo{year}{1993}) \bibinfo{pages}{3051--3054}.
  \DOIprefix\doi{10.1103/PhysRevLett.71.3051}.
  \href{http://arxiv.org/abs/hep-ph/9303313}{{\tt arXiv:hep-ph/9303313}}.
\bibitem[{Kolb and Tkachev(1996)}]{Kolb:1995bu}
\bibinfo{author}{E.~W. Kolb}, \bibinfo{author}{I.~I. Tkachev},
\newblock \bibinfo{title}{{Femtolensing and picolensing by axion
  miniclusters}},
\newblock \bibinfo{journal}{Astrophys. J. Lett.} \bibinfo{volume}{460}
  (\bibinfo{year}{1996}) \bibinfo{pages}{L25--L28}.
  \DOIprefix\doi{10.1086/309962}.
  \href{http://arxiv.org/abs/astro-ph/9510043}{{\tt arXiv:astro-ph/9510043}}.
\bibitem[{Guth et~al.(2015)Guth, Hertzberg, and
  Prescod-Weinstein}]{Guth:2014hsa}
\bibinfo{author}{A.~H. Guth}, \bibinfo{author}{M.~P. Hertzberg},
  \bibinfo{author}{C.~Prescod-Weinstein},
\newblock \bibinfo{title}{{Do Dark Matter Axions Form a Condensate with
  Long-Range Correlation?}},
\newblock \bibinfo{journal}{Phys. Rev. D} \bibinfo{volume}{92}
  (\bibinfo{year}{2015}) \bibinfo{pages}{103513}.
  \DOIprefix\doi{10.1103/PhysRevD.92.103513}.
  \href{http://arxiv.org/abs/1412.5930}{{\tt arXiv:1412.5930}}.
\bibitem[{Schiappacasse and Hertzberg(2018)}]{Schiappacasse:2017ham}
\bibinfo{author}{E.~D. Schiappacasse}, \bibinfo{author}{M.~P. Hertzberg},
\newblock \bibinfo{title}{{Analysis of Dark Matter Axion Clumps with Spherical
  Symmetry}},
\newblock \bibinfo{journal}{JCAP} \bibinfo{volume}{01} (\bibinfo{year}{2018})
  \bibinfo{pages}{037}. \DOIprefix\doi{10.1088/1475-7516/2018/01/037}.
  \href{http://arxiv.org/abs/1710.04729}{{\tt arXiv:1710.04729}},
  \bibinfo{note}{[Erratum: JCAP 03, E01 (2018)]}.
\bibitem[{Levkov et~al.(2018)Levkov, Panin, and Tkachev}]{Levkov:2018kau}
\bibinfo{author}{D.~G. Levkov}, \bibinfo{author}{A.~G. Panin},
  \bibinfo{author}{I.~I. Tkachev},
\newblock \bibinfo{title}{{Gravitational Bose-Einstein condensation in the
  kinetic regime}},
\newblock \bibinfo{journal}{Phys. Rev. Lett.} \bibinfo{volume}{121}
  (\bibinfo{year}{2018}) \bibinfo{pages}{151301}.
  \DOIprefix\doi{10.1103/PhysRevLett.121.151301}.
  \href{http://arxiv.org/abs/1804.05857}{{\tt arXiv:1804.05857}}.
\bibitem[{Visinelli et~al.(2018)Visinelli, Baum, Redondo, Freese, and
  Wilczek}]{Visinelli:2017ooc}
\bibinfo{author}{L.~Visinelli}, \bibinfo{author}{S.~Baum},
  \bibinfo{author}{J.~Redondo}, \bibinfo{author}{K.~Freese},
  \bibinfo{author}{F.~Wilczek},
\newblock \bibinfo{title}{{Dilute and dense axion stars}},
\newblock \bibinfo{journal}{Phys. Lett. B} \bibinfo{volume}{777}
  (\bibinfo{year}{2018}) \bibinfo{pages}{64--72}.
  \DOIprefix\doi{10.1016/j.physletb.2017.12.010}.
  \href{http://arxiv.org/abs/1710.08910}{{\tt arXiv:1710.08910}}.
\bibitem[{Seidel and Suen(1993)}]{Seidel.Suen.1993}
\bibinfo{author}{E.~Seidel}, \bibinfo{author}{W.-M. Suen},
\newblock \bibinfo{title}{{Formation of Solitonic Stars Through Gravitational
  Cooling}},
\newblock \bibinfo{journal}{arXiv}  (\bibinfo{year}{1993}).
  \DOIprefix\doi{10.1103/physrevlett.72.2516}.
  \href{http://arxiv.org/abs/gr-qc/9309015}{{\tt arXiv:gr-qc/9309015}}.
\bibitem[{Schive et~al.(2014)Schive, Chiueh, and
  Broadhurst}]{Schive.Broadhurst.2014}
\bibinfo{author}{H.-Y. Schive}, \bibinfo{author}{T.~Chiueh},
  \bibinfo{author}{T.~Broadhurst},
\newblock \bibinfo{title}{{Cosmic Structure as the Quantum Interference of a
  Coherent Dark Wave}},
\newblock \bibinfo{journal}{arXiv}  (\bibinfo{year}{2014}).
  \DOIprefix\doi{10.1038/nphys2996}. \href{http://arxiv.org/abs/1406.6586}{{\tt
  arXiv:1406.6586}}.
\bibitem[{Eggemeier and Niemeyer(2019)}]{Eggemeier.Niemeyer.2019}
\bibinfo{author}{B.~Eggemeier}, \bibinfo{author}{J.~C. Niemeyer},
\newblock \bibinfo{title}{{Formation and mass growth of axion stars in axion
  miniclusters}},
\newblock \bibinfo{journal}{arXiv}  (\bibinfo{year}{2019}).
  \DOIprefix\doi{10.1103/physrevd.100.063528}.
  \href{http://arxiv.org/abs/1906.01348}{{\tt arXiv:1906.01348}}.
\bibitem[{Chen et~al.(2020)Chen, Du, Lentz, Marsh, and
  Niemeyer}]{Chen.Niemeyer.2020}
\bibinfo{author}{J.~Chen}, \bibinfo{author}{X.~Du}, \bibinfo{author}{E.~W.
  Lentz}, \bibinfo{author}{D.~J.~E. Marsh}, \bibinfo{author}{J.~C. Niemeyer},
\newblock \bibinfo{title}{{New insights into the formation and growth of boson
  stars in dark matter halos}},
\newblock \bibinfo{journal}{arXiv}  (\bibinfo{year}{2020}).
  \DOIprefix\doi{10.1103/physrevd.104.083022}.
  \href{http://arxiv.org/abs/2011.01333}{{\tt arXiv:2011.01333}}.
\bibitem[{Croon et~al.(2020{\natexlab{a}})Croon, McKeen, Raj, and
  Wang}]{Croon:2020ouk}
\bibinfo{author}{D.~Croon}, \bibinfo{author}{D.~McKeen},
  \bibinfo{author}{N.~Raj}, \bibinfo{author}{Z.~Wang},
\newblock \bibinfo{title}{{Subaru-HSC through a different lens: Microlensing by
  extended dark matter structures}},
\newblock \bibinfo{journal}{Phys. Rev. D} \bibinfo{volume}{102}
  (\bibinfo{year}{2020}{\natexlab{a}}) \bibinfo{pages}{083021}.
  \DOIprefix\doi{10.1103/PhysRevD.102.083021}.
  \href{http://arxiv.org/abs/2007.12697}{{\tt arXiv:2007.12697}}.
\bibitem[{Croon et~al.(2020{\natexlab{b}})Croon, McKeen, and
  Raj}]{Croon:2020wpr}
\bibinfo{author}{D.~Croon}, \bibinfo{author}{D.~McKeen},
  \bibinfo{author}{N.~Raj},
\newblock \bibinfo{title}{{Gravitational microlensing by dark matter in
  extended structures}},
\newblock \bibinfo{journal}{Phys. Rev. D} \bibinfo{volume}{101}
  (\bibinfo{year}{2020}{\natexlab{b}}) \bibinfo{pages}{083013}.
  \DOIprefix\doi{10.1103/PhysRevD.101.083013}.
  \href{http://arxiv.org/abs/2002.08962}{{\tt arXiv:2002.08962}}.
\bibitem[{Fujikura et~al.(2021)Fujikura, Hertzberg, Schiappacasse, and
  Yamaguchi}]{Fujikura.Yamaguchi.2021}
\bibinfo{author}{K.~Fujikura}, \bibinfo{author}{M.~P. Hertzberg},
  \bibinfo{author}{E.~D. Schiappacasse}, \bibinfo{author}{M.~Yamaguchi},
\newblock \bibinfo{title}{{Microlensing constraints on axion stars including
  finite lens and source size effects}},
\newblock \bibinfo{journal}{arXiv}  (\bibinfo{year}{2021}).
  \DOIprefix\doi{10.1103/physrevd.104.123012}.
  \href{http://arxiv.org/abs/2109.04283}{{\tt arXiv:2109.04283}}.
\bibitem[{Niikura et~al.(2019{\natexlab{a}})}]{Niikura:2017zjd}
\bibinfo{author}{H.~Niikura}, et~al.,
\newblock \bibinfo{title}{{Microlensing constraints on primordial black holes
  with Subaru/HSC Andromeda observations}},
\newblock \bibinfo{journal}{Nat. Astron.} \bibinfo{volume}{3}
  (\bibinfo{year}{2019}{\natexlab{a}}) \bibinfo{pages}{524--534}.
  \DOIprefix\doi{10.1038/s41550-019-0723-1}.
  \href{http://arxiv.org/abs/1701.02151}{{\tt arXiv:1701.02151}}.
\bibitem[{Niikura et~al.(2019{\natexlab{b}})Niikura, Takada, Yokoyama, Sumi,
  and Masaki}]{Niikura:2019kqi}
\bibinfo{author}{H.~Niikura}, \bibinfo{author}{M.~Takada},
  \bibinfo{author}{S.~Yokoyama}, \bibinfo{author}{T.~Sumi},
  \bibinfo{author}{S.~Masaki},
\newblock \bibinfo{title}{{Constraints on Earth-mass primordial black holes
  from OGLE 5-year microlensing events}},
\newblock \bibinfo{journal}{Phys. Rev.} \bibinfo{volume}{D99}
  (\bibinfo{year}{2019}{\natexlab{b}}) \bibinfo{pages}{083503}.
  \DOIprefix\doi{10.1103/PhysRevD.99.083503}.
  \href{http://arxiv.org/abs/1901.07120}{{\tt arXiv:1901.07120}}.
\bibitem[{Grilli~di Cortona et~al.(2016)Grilli~di Cortona, Hardy, Pardo~Vega,
  and Villadoro}]{GrillidiCortona:2015jxo}
\bibinfo{author}{G.~Grilli~di Cortona}, \bibinfo{author}{E.~Hardy},
  \bibinfo{author}{J.~Pardo~Vega}, \bibinfo{author}{G.~Villadoro},
\newblock \bibinfo{title}{{The QCD axion, precisely}},
\newblock \bibinfo{journal}{JHEP} \bibinfo{volume}{01} (\bibinfo{year}{2016})
  \bibinfo{pages}{034}. \DOIprefix\doi{10.1007/JHEP01(2016)034}.
  \href{http://arxiv.org/abs/1511.02867}{{\tt arXiv:1511.02867}}.
\bibitem[{Levkov et~al.(2017)Levkov, Panin, and Tkachev}]{Levkov:2016rkk}
\bibinfo{author}{D.~G. Levkov}, \bibinfo{author}{A.~G. Panin},
  \bibinfo{author}{I.~I. Tkachev},
\newblock \bibinfo{title}{{Relativistic axions from collapsing Bose stars}},
\newblock \bibinfo{journal}{Phys. Rev. Lett.} \bibinfo{volume}{118}
  (\bibinfo{year}{2017}) \bibinfo{pages}{011301}.
  \DOIprefix\doi{10.1103/PhysRevLett.118.011301}.
  \href{http://arxiv.org/abs/1609.03611}{{\tt arXiv:1609.03611}}.
\bibitem[{Eby et~al.(2021)Eby, Shirai, Stadnik, and Takhistov}]{Eby:2021ece}
\bibinfo{author}{J.~Eby}, \bibinfo{author}{S.~Shirai}, \bibinfo{author}{Y.~V.
  Stadnik}, \bibinfo{author}{V.~Takhistov},
\newblock \bibinfo{title}{{Probing Relativistic Axions from Transient
  Astrophysical Sources}}  (\bibinfo{year}{2021}).
  \href{http://arxiv.org/abs/2106.14893}{{\tt arXiv:2106.14893}}.
\bibitem[{Helfer et~al.(2017)Helfer, Marsh, Clough, Fairbairn, Lim, and
  Becerril}]{Helfer.Becerril.2017}
\bibinfo{author}{T.~Helfer}, \bibinfo{author}{D.~J.~E. Marsh},
  \bibinfo{author}{K.~Clough}, \bibinfo{author}{M.~Fairbairn},
  \bibinfo{author}{E.~A. Lim}, \bibinfo{author}{R.~Becerril},
\newblock \bibinfo{title}{{Black hole formation from axion stars}},
\newblock \bibinfo{journal}{Journal of Cosmology and Astroparticle Physics}
  \bibinfo{volume}{2017} (\bibinfo{year}{2017}) \bibinfo{pages}{055--055}.
  \DOIprefix\doi{10.1088/1475-7516/2017/03/055}.
  \href{http://arxiv.org/abs/1609.04724}{{\tt arXiv:1609.04724}}.
\bibitem[{Michel and Moss(2018)}]{Michel.Moss.2018}
\bibinfo{author}{F.~Michel}, \bibinfo{author}{I.~G. Moss},
\newblock \bibinfo{title}{{Relativistic collapse of axion stars}},
\newblock \bibinfo{journal}{Physics Letters B} \bibinfo{volume}{785}
  (\bibinfo{year}{2018}) \bibinfo{pages}{9--13}.
  \DOIprefix\doi{10.1016/j.physletb.2018.07.063}.
  \href{http://arxiv.org/abs/1802.10085}{{\tt arXiv:1802.10085}}.
\bibitem[{Linde(1982)}]{Linde:1982uu}
\bibinfo{author}{A.~D. Linde},
\newblock \bibinfo{title}{{Scalar Field Fluctuations in Expanding Universe and
  the New Inflationary Universe Scenario}},
\newblock \bibinfo{journal}{Phys. Lett.} \bibinfo{volume}{B116}
  (\bibinfo{year}{1982}) \bibinfo{pages}{335--339}.
  \DOIprefix\doi{10.1016/0370-2693(82)90293-3}.
\bibitem[{Wilczek(2004)}]{Wilczek:2004cr}
\bibinfo{author}{F.~Wilczek},
\newblock \bibinfo{title}{{A Model of anthropic reasoning, addressing the dark
  to ordinary matter coincidence}}  (\bibinfo{year}{2004}).
  \href{http://arxiv.org/abs/hep-ph/0408167}{{\tt arXiv:hep-ph/0408167}}.
\bibitem[{Tegmark et~al.(2006)Tegmark, Aguirre, Rees, and
  Wilczek}]{Tegmark:2005dy}
\bibinfo{author}{M.~Tegmark}, \bibinfo{author}{A.~Aguirre},
  \bibinfo{author}{M.~Rees}, \bibinfo{author}{F.~Wilczek},
\newblock \bibinfo{title}{{Dimensionless constants, cosmology and other dark
  matters}},
\newblock \bibinfo{journal}{Phys. Rev. D} \bibinfo{volume}{73}
  (\bibinfo{year}{2006}) \bibinfo{pages}{023505}.
  \DOIprefix\doi{10.1103/PhysRevD.73.023505}.
  \href{http://arxiv.org/abs/astro-ph/0511774}{{\tt arXiv:astro-ph/0511774}}.
\bibitem[{Kusenko et~al.(2013)Kusenko, Loewenstein, and
  Yanagida}]{Kusenko:2012ch}
\bibinfo{author}{A.~Kusenko}, \bibinfo{author}{M.~Loewenstein},
  \bibinfo{author}{T.~T. Yanagida},
\newblock \bibinfo{title}{{Moduli dark matter and the search for its decay line
  using Suzaku X-ray telescope}},
\newblock \bibinfo{journal}{Phys. Rev. D} \bibinfo{volume}{87}
  (\bibinfo{year}{2013}) \bibinfo{pages}{043508}.
  \DOIprefix\doi{10.1103/PhysRevD.87.043508}.
  \href{http://arxiv.org/abs/1209.6403}{{\tt arXiv:1209.6403}}.
\bibitem[{Fuller et~al.(2011)Fuller, Kishimoto, and Kusenko}]{Fuller:2011qy}
\bibinfo{author}{G.~M. Fuller}, \bibinfo{author}{C.~T. Kishimoto},
  \bibinfo{author}{A.~Kusenko},
\newblock \bibinfo{title}{{Heavy sterile neutrinos, entropy and relativistic
  energy production, and the relic neutrino background}}
  (\bibinfo{year}{2011}). \href{http://arxiv.org/abs/1110.6479}{{\tt
  arXiv:1110.6479}}.
\bibitem[{Patwardhan et~al.(2015)Patwardhan, Fuller, Kishimoto, and
  Kusenko}]{Patwardhan:2015kga}
\bibinfo{author}{A.~V. Patwardhan}, \bibinfo{author}{G.~M. Fuller},
  \bibinfo{author}{C.~T. Kishimoto}, \bibinfo{author}{A.~Kusenko},
\newblock \bibinfo{title}{{Diluted equilibrium sterile neutrino dark matter}},
\newblock \bibinfo{journal}{Phys. Rev. D} \bibinfo{volume}{92}
  (\bibinfo{year}{2015}) \bibinfo{pages}{103509}.
  \DOIprefix\doi{10.1103/PhysRevD.92.103509}.
  \href{http://arxiv.org/abs/1507.01977}{{\tt arXiv:1507.01977}}.
\bibitem[{Hasegawa et~al.(2019)Hasegawa, Hiroshima, Kohri, Hansen, Tram, and
  Hannestad}]{Hasegawa:2019jsa}
\bibinfo{author}{T.~Hasegawa}, \bibinfo{author}{N.~Hiroshima},
  \bibinfo{author}{K.~Kohri}, \bibinfo{author}{R.~S.~L. Hansen},
  \bibinfo{author}{T.~Tram}, \bibinfo{author}{S.~Hannestad},
\newblock \bibinfo{title}{{MeV-scale reheating temperature and thermalization
  of oscillating neutrinos by radiative and hadronic decays of massive
  particles}},
\newblock \bibinfo{journal}{JCAP} \bibinfo{volume}{12} (\bibinfo{year}{2019})
  \bibinfo{pages}{012}. \DOIprefix\doi{10.1088/1475-7516/2019/12/012}.
  \href{http://arxiv.org/abs/1908.10189}{{\tt arXiv:1908.10189}}.
\bibitem[{Axenides et~al.(1983)Axenides, Brandenberger, and
  Turner}]{AXENIDES1983178}
\bibinfo{author}{M.~Axenides}, \bibinfo{author}{R.~Brandenberger},
  \bibinfo{author}{M.~Turner},
\newblock \bibinfo{title}{Development of axion perturbations in an axion
  dominated universe},
\newblock \bibinfo{journal}{Physics Letters B} \bibinfo{volume}{126}
  (\bibinfo{year}{1983}) \bibinfo{pages}{178--182}. \URLprefix
  \url{https://www.sciencedirect.com/science/article/pii/0370269383905865}.
  \DOIprefix\doi{https://doi.org/10.1016/0370-2693(83)90586-5}.
\bibitem[{Linde(1985)}]{LINDE1985375}
\bibinfo{author}{A.~Linde},
\newblock \bibinfo{title}{Generation of isothermal density perturbations in the
  inflationary universe},
\newblock \bibinfo{journal}{Physics Letters B} \bibinfo{volume}{158}
  (\bibinfo{year}{1985}) \bibinfo{pages}{375--380}. \URLprefix
  \url{https://www.sciencedirect.com/science/article/pii/0370269385904368}.
  \DOIprefix\doi{https://doi.org/10.1016/0370-2693(85)90436-8}.
\bibitem[{Lyth(1990)}]{LYTH1990408}
\bibinfo{author}{D.~H. Lyth},
\newblock \bibinfo{title}{A limit on the inflationary energy density from axion
  isocurvature fluctuations},
\newblock \bibinfo{journal}{Physics Letters B} \bibinfo{volume}{236}
  (\bibinfo{year}{1990}) \bibinfo{pages}{408--410}. \URLprefix
  \url{https://www.sciencedirect.com/science/article/pii/037026939090374F}.
  \DOIprefix\doi{https://doi.org/10.1016/0370-2693(90)90374-F}.
\bibitem[{Linde and Lyth(1990)}]{LINDE1990353}
\bibinfo{author}{A.~D. Linde}, \bibinfo{author}{D.~H. Lyth},
\newblock \bibinfo{title}{Axionic domain wall production during inflation},
\newblock \bibinfo{journal}{Physics Letters B} \bibinfo{volume}{246}
  (\bibinfo{year}{1990}) \bibinfo{pages}{353--358}. \URLprefix
  \url{https://www.sciencedirect.com/science/article/pii/037026939090613B}.
  \DOIprefix\doi{https://doi.org/10.1016/0370-2693(90)90613-B}.
\bibitem[{Turner and Wilczek(1991)}]{Turner:1990uz}
\bibinfo{author}{M.~S. Turner}, \bibinfo{author}{F.~Wilczek},
\newblock \bibinfo{title}{{Inflationary axion cosmology}},
\newblock \bibinfo{journal}{Phys. Rev. Lett.} \bibinfo{volume}{66}
  (\bibinfo{year}{1991}) \bibinfo{pages}{5--8}.
  \DOIprefix\doi{10.1103/PhysRevLett.66.5}.
\bibitem[{Seckel and Turner(1985)}]{PhysRevD.32.3178}
\bibinfo{author}{D.~Seckel}, \bibinfo{author}{M.~S. Turner},
\newblock \bibinfo{title}{``isothermal'' density perturbations in an
  axion-dominated inflationary universe},
\newblock \bibinfo{journal}{Phys. Rev. D} \bibinfo{volume}{32}
  (\bibinfo{year}{1985}) \bibinfo{pages}{3178--3183}. \URLprefix
  \url{https://link.aps.org/doi/10.1103/PhysRevD.32.3178}.
  \DOIprefix\doi{10.1103/PhysRevD.32.3178}.
\bibitem[{{Linde}(1991)}]{1991PhLB..259...38L}
\bibinfo{author}{A.~{Linde}},
\newblock \bibinfo{title}{{Axions in inflationary cosmology}},
\newblock \bibinfo{journal}{Physics Letters B} \bibinfo{volume}{259}
  (\bibinfo{year}{1991}) \bibinfo{pages}{38--47}.
  \DOIprefix\doi{10.1016/0370-2693(91)90130-I}.
\bibitem[{Visinelli and Gondolo(2010)}]{Visinelli:2009kt}
\bibinfo{author}{L.~Visinelli}, \bibinfo{author}{P.~Gondolo},
\newblock \bibinfo{title}{{Axion cold dark matter in non-standard
  cosmologies}},
\newblock \bibinfo{journal}{Phys. Rev. D} \bibinfo{volume}{81}
  (\bibinfo{year}{2010}) \bibinfo{pages}{063508}.
  \DOIprefix\doi{10.1103/PhysRevD.81.063508}.
  \href{http://arxiv.org/abs/0912.0015}{{\tt arXiv:0912.0015}}.
\bibitem[{Kawasaki et~al.(1996)Kawasaki, Moroi, and Yanagida}]{Kawasaki:1995vt}
\bibinfo{author}{M.~Kawasaki}, \bibinfo{author}{T.~Moroi},
  \bibinfo{author}{T.~Yanagida},
\newblock \bibinfo{title}{{Can decaying particles raise the upper bound on the
  Peccei-Quinn scale?}},
\newblock \bibinfo{journal}{Phys. Lett. B} \bibinfo{volume}{383}
  (\bibinfo{year}{1996}) \bibinfo{pages}{313--316}.
  \DOIprefix\doi{10.1016/0370-2693(96)00743-5}.
  \href{http://arxiv.org/abs/hep-ph/9510461}{{\tt arXiv:hep-ph/9510461}}.
\bibitem[{Kobayashi et~al.(2013)Kobayashi, Kurematsu, and
  Takahashi}]{Kobayashi:2013nva}
\bibinfo{author}{T.~Kobayashi}, \bibinfo{author}{R.~Kurematsu},
  \bibinfo{author}{F.~Takahashi},
\newblock \bibinfo{title}{{Isocurvature Constraints and Anharmonic Effects on
  QCD Axion Dark Matter}},
\newblock \bibinfo{journal}{JCAP} \bibinfo{volume}{09} (\bibinfo{year}{2013})
  \bibinfo{pages}{032}. \DOIprefix\doi{10.1088/1475-7516/2013/09/032}.
  \href{http://arxiv.org/abs/1304.0922}{{\tt arXiv:1304.0922}}.
\bibitem[{Takahashi et~al.(2018)Takahashi, Yin, and Guth}]{Takahashi:2018tdu}
\bibinfo{author}{F.~Takahashi}, \bibinfo{author}{W.~Yin},
  \bibinfo{author}{A.~H. Guth},
\newblock \bibinfo{title}{{QCD axion window and low-scale inflation}},
\newblock \bibinfo{journal}{Phys. Rev. D} \bibinfo{volume}{98}
  (\bibinfo{year}{2018}) \bibinfo{pages}{015042}.
  \DOIprefix\doi{10.1103/PhysRevD.98.015042}.
  \href{http://arxiv.org/abs/1805.08763}{{\tt arXiv:1805.08763}}.
\bibitem[{Schiappacasse and Yanagida(2021)}]{Schiappacasse:2021zlr}
\bibinfo{author}{E.~D. Schiappacasse}, \bibinfo{author}{T.~T. Yanagida},
\newblock \bibinfo{title}{{Can QCD axion stars explain Subaru HSC
  microlensing?}},
\newblock \bibinfo{journal}{Phys. Rev. D} \bibinfo{volume}{104}
  (\bibinfo{year}{2021}) \bibinfo{pages}{103020}.
  \DOIprefix\doi{10.1103/PhysRevD.104.103020}.
  \href{http://arxiv.org/abs/2109.13153}{{\tt arXiv:2109.13153}}.
\bibitem[{{Paczynski}(1986)}]{1986ApJ...304....1P}
\bibinfo{author}{B.~{Paczynski}},
\newblock \bibinfo{title}{{Gravitational Microlensing by the Galactic Halo}},
\newblock \bibinfo{journal}{\apj} \bibinfo{volume}{304} (\bibinfo{year}{1986})
  \bibinfo{pages}{1}. \DOIprefix\doi{10.1086/164140}.
\bibitem[{Inomata et~al.(2017)Inomata, Kawasaki, Mukaida, Tada, and
  Yanagida}]{Inomata:2017okj}
\bibinfo{author}{K.~Inomata}, \bibinfo{author}{M.~Kawasaki},
  \bibinfo{author}{K.~Mukaida}, \bibinfo{author}{Y.~Tada},
  \bibinfo{author}{T.~T. Yanagida},
\newblock \bibinfo{title}{{Inflationary Primordial Black Holes as All Dark
  Matter}},
\newblock \bibinfo{journal}{Phys. Rev. D} \bibinfo{volume}{96}
  (\bibinfo{year}{2017}) \bibinfo{pages}{043504}.
  \DOIprefix\doi{10.1103/PhysRevD.96.043504}.
  \href{http://arxiv.org/abs/1701.02544}{{\tt arXiv:1701.02544}}.
\bibitem[{{Mr{\'o}z} et~al.(2017){Mr{\'o}z}, {Udalski}, {Skowron}, {Poleski},
  {Koz{\l}owski}, {Szyma{\'n}ski}, {Soszy{\'n}ski}, {Wyrzykowski},
  {Pietrukowicz}, {Ulaczyk}, {Skowron}, and {Pawlak}}]{2017Natur.548..183M}
\bibinfo{author}{P.~{Mr{\'o}z}}, \bibinfo{author}{A.~{Udalski}},
  \bibinfo{author}{J.~{Skowron}}, \bibinfo{author}{R.~{Poleski}},
  \bibinfo{author}{S.~{Koz{\l}owski}}, \bibinfo{author}{M.~K. {Szyma{\'n}ski}},
  \bibinfo{author}{I.~{Soszy{\'n}ski}}, \bibinfo{author}{{\L}.~{Wyrzykowski}},
  \bibinfo{author}{P.~{Pietrukowicz}}, \bibinfo{author}{K.~{Ulaczyk}},
  \bibinfo{author}{D.~{Skowron}}, \bibinfo{author}{M.~{Pawlak}},
\newblock \bibinfo{title}{{No large population of unbound or wide-orbit
  Jupiter-mass planets}},
\newblock \bibinfo{journal}{\nat} \bibinfo{volume}{548} (\bibinfo{year}{2017})
  \bibinfo{pages}{183--186}. \DOIprefix\doi{10.1038/nature23276}.
  \href{http://arxiv.org/abs/1707.07634}{{\tt arXiv:1707.07634}}.
\bibitem[{Sugiyama et~al.(2020)Sugiyama, Kurita, and
  Takada}]{Sugiyama.Takada.2020}
\bibinfo{author}{S.~Sugiyama}, \bibinfo{author}{T.~Kurita},
  \bibinfo{author}{M.~Takada},
\newblock \bibinfo{title}{{On the wave optics effect on primordial black hole
  constraints from optical microlensing search}},
\newblock \bibinfo{journal}{Monthly Notices of the Royal Astronomical Society}
  \bibinfo{volume}{493} (\bibinfo{year}{2020}) \bibinfo{pages}{3632--3641}.
  \DOIprefix\doi{10.1093/mnras/staa407}.
  \href{http://arxiv.org/abs/1905.06066}{{\tt arXiv:1905.06066}}.
\bibitem[{{Sumi} et~al.(2003){Sumi}, {Abe}, {Bond}, {Dodd}, {Hearnshaw},
  {Honda}, {Honma}, {Kan-ya}, {Kilmartin}, {Masuda}, {Matsubara}, {Muraki},
  {Nakamura}, {Nishi}, {Noda}, {Ohnishi}, {Petterson}, {Rattenbury}, {Reid},
  {Saito}, {Saito}, {Sato}, {Sekiguchi}, {Skuljan}, {Sullivan}, {Takeuti},
  {Tristram}, {Wilkinson}, {Yanagisawa}, and {Yock}}]{2003ApJ...591..204S}
\bibinfo{author}{T.~{Sumi}}, \bibinfo{author}{F.~{Abe}}, \bibinfo{author}{I.~A.
  {Bond}}, \bibinfo{author}{R.~J. {Dodd}}, \bibinfo{author}{J.~B. {Hearnshaw}},
  \bibinfo{author}{M.~{Honda}}, \bibinfo{author}{M.~{Honma}},
  \bibinfo{author}{Y.~{Kan-ya}}, \bibinfo{author}{P.~M. {Kilmartin}},
  \bibinfo{author}{K.~{Masuda}}, \bibinfo{author}{Y.~{Matsubara}},
  \bibinfo{author}{Y.~{Muraki}}, \bibinfo{author}{T.~{Nakamura}},
  \bibinfo{author}{R.~{Nishi}}, \bibinfo{author}{S.~{Noda}},
  \bibinfo{author}{K.~{Ohnishi}}, \bibinfo{author}{O.~K.~L. {Petterson}},
  \bibinfo{author}{N.~J. {Rattenbury}}, \bibinfo{author}{M.~{Reid}},
  \bibinfo{author}{T.~{Saito}}, \bibinfo{author}{Y.~{Saito}},
  \bibinfo{author}{H.~{Sato}}, \bibinfo{author}{M.~{Sekiguchi}},
  \bibinfo{author}{J.~{Skuljan}}, \bibinfo{author}{D.~J. {Sullivan}},
  \bibinfo{author}{M.~{Takeuti}}, \bibinfo{author}{P.~J. {Tristram}},
  \bibinfo{author}{S.~{Wilkinson}}, \bibinfo{author}{T.~{Yanagisawa}},
  \bibinfo{author}{P.~C.~M. {Yock}},
\newblock \bibinfo{title}{{Microlensing Optical Depth toward the Galactic Bulge
  from Microlensing Observations in Astrophysics Group Observations during 2000
  with Difference Image Analysis}},
\newblock \bibinfo{journal}{\apj} \bibinfo{volume}{591} (\bibinfo{year}{2003})
  \bibinfo{pages}{204--227}. \DOIprefix\doi{10.1086/375212}.
  \href{http://arxiv.org/abs/astro-ph/0207604}{{\tt arXiv:astro-ph/0207604}}.
\bibitem[{{Sumi} et~al.(2011){Sumi}, {Kamiya}, {Bennett}, {Bond}, {Abe},
  {Botzler}, {Fukui}, {Furusawa}, {Hearnshaw}, {Itow}, {Kilmartin}, {Korpela},
  {Lin}, {Ling}, {Masuda}, {Matsubara}, {Miyake}, {Motomura}, {Muraki},
  {Nagaya}, {Nakamura}, {Ohnishi}, {Okumura}, {Perrott}, {Rattenbury}, {Saito},
  {Sako}, {Sullivan}, {Sweatman}, {Tristram}, {Udalski}, {Szyma{\'n}ski},
  {Kubiak}, {Pietrzy{\'n}ski}, {Poleski}, {Soszy{\'n}ski}, {Wyrzykowski},
  {Ulaczyk}, and {Microlensing Observations in Astrophysics (MOA)
  Collaboration}}]{2011Natur.473..349S}
\bibinfo{author}{T.~{Sumi}}, \bibinfo{author}{K.~{Kamiya}},
  \bibinfo{author}{D.~P. {Bennett}}, \bibinfo{author}{I.~A. {Bond}},
  \bibinfo{author}{F.~{Abe}}, \bibinfo{author}{C.~S. {Botzler}},
  \bibinfo{author}{A.~{Fukui}}, \bibinfo{author}{K.~{Furusawa}},
  \bibinfo{author}{J.~B. {Hearnshaw}}, \bibinfo{author}{Y.~{Itow}},
  \bibinfo{author}{P.~M. {Kilmartin}}, \bibinfo{author}{A.~{Korpela}},
  \bibinfo{author}{W.~{Lin}}, \bibinfo{author}{C.~H. {Ling}},
  \bibinfo{author}{K.~{Masuda}}, \bibinfo{author}{Y.~{Matsubara}},
  \bibinfo{author}{N.~{Miyake}}, \bibinfo{author}{M.~{Motomura}},
  \bibinfo{author}{Y.~{Muraki}}, \bibinfo{author}{M.~{Nagaya}},
  \bibinfo{author}{S.~{Nakamura}}, \bibinfo{author}{K.~{Ohnishi}},
  \bibinfo{author}{T.~{Okumura}}, \bibinfo{author}{Y.~C. {Perrott}},
  \bibinfo{author}{N.~{Rattenbury}}, \bibinfo{author}{T.~{Saito}},
  \bibinfo{author}{T.~{Sako}}, \bibinfo{author}{D.~J. {Sullivan}},
  \bibinfo{author}{W.~L. {Sweatman}}, \bibinfo{author}{P.~J. {Tristram}},
  \bibinfo{author}{A.~{Udalski}}, \bibinfo{author}{M.~K. {Szyma{\'n}ski}},
  \bibinfo{author}{M.~{Kubiak}}, \bibinfo{author}{G.~{Pietrzy{\'n}ski}},
  \bibinfo{author}{R.~{Poleski}}, \bibinfo{author}{I.~{Soszy{\'n}ski}},
  \bibinfo{author}{{\L}.~{Wyrzykowski}}, \bibinfo{author}{K.~{Ulaczyk}},
  \bibinfo{author}{{Microlensing Observations in Astrophysics (MOA)
  Collaboration}},
\newblock \bibinfo{title}{{Unbound or distant planetary mass population
  detected by gravitational microlensing}},
\newblock \bibinfo{journal}{\nat} \bibinfo{volume}{473} (\bibinfo{year}{2011})
  \bibinfo{pages}{349--352}. \DOIprefix\doi{10.1038/nature10092}.
  \href{http://arxiv.org/abs/1105.3544}{{\tt arXiv:1105.3544}}.
\bibitem[{Abrams and Takada(2020)}]{Abrams:2020jvs}
\bibinfo{author}{N.~S. Abrams}, \bibinfo{author}{M.~Takada},
\newblock \bibinfo{title}{{Hunting gravitational wave black holes with
  microlensing}}  (\bibinfo{year}{2020}).
  \href{http://arxiv.org/abs/2006.05578}{{\tt arXiv:2006.05578}}.
\bibitem[{{Toki} and {Takada}(2021)}]{2021arXiv210313015T}
\bibinfo{author}{S.~{Toki}}, \bibinfo{author}{M.~{Takada}},
\newblock \bibinfo{title}{{Finding gravitational-wave black holes with parallax
  microlensing}},
\newblock \bibinfo{journal}{arXiv e-prints}  (\bibinfo{year}{2021})
  \bibinfo{pages}{arXiv:2103.13015}.
  \href{http://arxiv.org/abs/2103.13015}{{\tt arXiv:2103.13015}}.
\bibitem[{{Navarro} et~al.(1997){Navarro}, {Frenk}, and
  {White}}]{1997ApJ...490..493N}
\bibinfo{author}{J.~F. {Navarro}}, \bibinfo{author}{C.~S. {Frenk}},
  \bibinfo{author}{S.~D.~M. {White}},
\newblock \bibinfo{title}{{A Universal Density Profile from Hierarchical
  Clustering}},
\newblock \bibinfo{journal}{\apj} \bibinfo{volume}{490} (\bibinfo{year}{1997})
  \bibinfo{pages}{493--508}. \DOIprefix\doi{10.1086/304888}.
  \href{http://arxiv.org/abs/astro-ph/9611107}{{\tt arXiv:astro-ph/9611107}}.
\bibitem[{{Klypin} et~al.(2002){Klypin}, {Zhao}, and
  {Somerville}}]{Klypinetal:02}
\bibinfo{author}{A.~{Klypin}}, \bibinfo{author}{H.~{Zhao}},
  \bibinfo{author}{R.~S. {Somerville}},
\newblock \bibinfo{title}{{{$\Lambda$}CDM-based Models for the Milky Way and
  M31. I. Dynamical Models}},
\newblock \bibinfo{journal}{\apj} \bibinfo{volume}{573} (\bibinfo{year}{2002})
  \bibinfo{pages}{597--613}. \DOIprefix\doi{10.1086/340656}.
  \href{http://arxiv.org/abs/astro-ph/0110390}{{\tt arXiv:astro-ph/0110390}}.
\bibitem[{Di~Luzio et~al.(2017)Di~Luzio, Mescia, and Nardi}]{DiLuzio:2016sbl}
\bibinfo{author}{L.~Di~Luzio}, \bibinfo{author}{F.~Mescia},
  \bibinfo{author}{E.~Nardi},
\newblock \bibinfo{title}{{Redefining the Axion Window}},
\newblock \bibinfo{journal}{Phys. Rev. Lett.} \bibinfo{volume}{118}
  (\bibinfo{year}{2017}) \bibinfo{pages}{031801}.
  \DOIprefix\doi{10.1103/PhysRevLett.118.031801}.
  \href{http://arxiv.org/abs/1610.07593}{{\tt arXiv:1610.07593}}.
\bibitem[{{Raffelt}(2008)}]{2008LNP...741...51R}
\bibinfo{author}{G.~G. {Raffelt}}, \bibinfo{title}{{Astrophysical Axion
  Bounds}}, volume \bibinfo{volume}{741}, \bibinfo{year}{2008},
  p.~\bibinfo{pages}{51}.
\bibitem[{Abramowski et~al.(2013)}]{HESS:2013udx}
\bibinfo{author}{A.~Abramowski}, et~al. (\bibinfo{collaboration}{H.E.S.S.}),
\newblock \bibinfo{title}{{Constraints on axionlike particles with H.E.S.S.
  from the irregularity of the PKS 2155-304 energy spectrum}},
\newblock \bibinfo{journal}{Phys. Rev. D} \bibinfo{volume}{88}
  (\bibinfo{year}{2013}) \bibinfo{pages}{102003}.
  \DOIprefix\doi{10.1103/PhysRevD.88.102003}.
  \href{http://arxiv.org/abs/1311.3148}{{\tt arXiv:1311.3148}}.
\bibitem[{{Shokair} et~al.(2014){Shokair}, {Root}, {van Bibber}, {Brubaker},
  {Gurevich}, {Cahn}, {Lamoreaux}, {Anil}, {Lehnert}, {Mitchell}, {Reed}, and
  {Carosi}}]{2014IJMPA..2943004S}
\bibinfo{author}{T.~M. {Shokair}}, \bibinfo{author}{J.~{Root}},
  \bibinfo{author}{K.~A. {van Bibber}}, \bibinfo{author}{B.~{Brubaker}},
  \bibinfo{author}{Y.~V. {Gurevich}}, \bibinfo{author}{S.~B. {Cahn}},
  \bibinfo{author}{S.~K. {Lamoreaux}}, \bibinfo{author}{M.~A. {Anil}},
  \bibinfo{author}{K.~W. {Lehnert}}, \bibinfo{author}{B.~K. {Mitchell}},
  \bibinfo{author}{A.~{Reed}}, \bibinfo{author}{G.~{Carosi}},
\newblock \bibinfo{title}{{Future directions in the microwave cavity search for
  dark matter axions}},
\newblock \bibinfo{journal}{International Journal of Modern Physics A}
  \bibinfo{volume}{29} (\bibinfo{year}{2014}) \bibinfo{pages}{1443004}.
  \DOIprefix\doi{10.1142/S0217751X14430040}.
  \href{http://arxiv.org/abs/1405.3685}{{\tt arXiv:1405.3685}}.
\bibitem[{{Kuster} et~al.(2007){Kuster}, {Br{\"a}uninger}, {Cebri{\'a}n},
  {Davenport}, {Eleftheriadis}, {Englhauser}, {Fischer}, {Franz}, {Friedrich},
  {Hartmann}, {Heinsius}, {Hoffmann}, {Hoffmeister}, {Joux}, {Kang},
  {K{\"o}nigsmann}, {Kotthaus}, {Papaevangelou}, {Lasseur}, {Lippitsch},
  {Lutz}, {Morales}, {Rodr{\'\i}guez}, {Str{\"u}der}, {Vogel}, and
  {Zioutas}}]{2007NJPh....9..169K}
\bibinfo{author}{M.~{Kuster}}, \bibinfo{author}{H.~{Br{\"a}uninger}},
  \bibinfo{author}{S.~{Cebri{\'a}n}}, \bibinfo{author}{M.~{Davenport}},
  \bibinfo{author}{C.~{Eleftheriadis}}, \bibinfo{author}{J.~{Englhauser}},
  \bibinfo{author}{H.~{Fischer}}, \bibinfo{author}{J.~{Franz}},
  \bibinfo{author}{P.~{Friedrich}}, \bibinfo{author}{R.~{Hartmann}},
  \bibinfo{author}{F.~H. {Heinsius}}, \bibinfo{author}{D.~H.~H. {Hoffmann}},
  \bibinfo{author}{G.~{Hoffmeister}}, \bibinfo{author}{J.~N. {Joux}},
  \bibinfo{author}{D.~{Kang}}, \bibinfo{author}{K.~{K{\"o}nigsmann}},
  \bibinfo{author}{R.~{Kotthaus}}, \bibinfo{author}{T.~{Papaevangelou}},
  \bibinfo{author}{C.~{Lasseur}}, \bibinfo{author}{A.~{Lippitsch}},
  \bibinfo{author}{G.~{Lutz}}, \bibinfo{author}{J.~{Morales}},
  \bibinfo{author}{A.~{Rodr{\'\i}guez}}, \bibinfo{author}{L.~{Str{\"u}der}},
  \bibinfo{author}{J.~{Vogel}}, \bibinfo{author}{{Zioutas}},
\newblock \bibinfo{title}{{The x-ray telescope of CAST}},
\newblock \bibinfo{journal}{New Journal of Physics} \bibinfo{volume}{9}
  (\bibinfo{year}{2007}) \bibinfo{pages}{169}.
  \DOIprefix\doi{10.1088/1367-2630/9/6/169}.
  \href{http://arxiv.org/abs/physics/0702188}{{\tt arXiv:physics/0702188}}.
\bibitem[{Anastassopoulos et~al.(2017)}]{CAST:2017uph}
\bibinfo{author}{V.~Anastassopoulos}, et~al. (\bibinfo{collaboration}{CAST}),
\newblock \bibinfo{title}{{New CAST Limit on the Axion-Photon Interaction}},
\newblock \bibinfo{journal}{Nature Phys.} \bibinfo{volume}{13}
  (\bibinfo{year}{2017}) \bibinfo{pages}{584--590}.
  \DOIprefix\doi{10.1038/nphys4109}.
  \href{http://arxiv.org/abs/1705.02290}{{\tt arXiv:1705.02290}}.
\bibitem[{{Kusenko} et~al.(2020){Kusenko}, {Sasaki}, {Sugiyama}, {Takada},
  {Takhistov}, and {Vitagliano}}]{2020PhRvL.125r1304K}
\bibinfo{author}{A.~{Kusenko}}, \bibinfo{author}{M.~{Sasaki}},
  \bibinfo{author}{S.~{Sugiyama}}, \bibinfo{author}{M.~{Takada}},
  \bibinfo{author}{V.~{Takhistov}}, \bibinfo{author}{E.~{Vitagliano}},
\newblock \bibinfo{title}{{Exploring Primordial Black Holes from the Multiverse
  with Optical Telescopes}},
\newblock \bibinfo{journal}{\prl} \bibinfo{volume}{125} (\bibinfo{year}{2020})
  \bibinfo{pages}{181304}. \DOIprefix\doi{10.1103/PhysRevLett.125.181304}.
  \href{http://arxiv.org/abs/2001.09160}{{\tt arXiv:2001.09160}}.
\bibitem[{Kimball et~al.(2018)Kimball, Budker, Eby, Pospelov, Pustelny,
  Scholtes, Stadnik, Weis, and Wickenbrock}]{Kimball.Wickenbrock.2018}
\bibinfo{author}{D.~F.~J. Kimball}, \bibinfo{author}{D.~Budker},
  \bibinfo{author}{J.~Eby}, \bibinfo{author}{M.~Pospelov},
  \bibinfo{author}{S.~Pustelny}, \bibinfo{author}{T.~Scholtes},
  \bibinfo{author}{Y.~V. Stadnik}, \bibinfo{author}{A.~Weis},
  \bibinfo{author}{A.~Wickenbrock},
\newblock \bibinfo{title}{{Searching for axion stars and Q-balls with a
  terrestrial magnetometer network}},
\newblock \bibinfo{journal}{Physical Review D} \bibinfo{volume}{97}
  (\bibinfo{year}{2018}) \bibinfo{pages}{043002}.
  \DOIprefix\doi{10.1103/physrevd.97.043002}.
  \href{http://arxiv.org/abs/1710.04323}{{\tt arXiv:1710.04323}}.
\bibitem[{Pustelny et~al.(2013)Pustelny, Kimball, Pankow, Ledbetter,
  Wlodarczyk, Wcislo, Pospelov, Smith, Read, Gawlik, and
  Budker}]{Pustelny.Budker.2013}
\bibinfo{author}{S.~Pustelny}, \bibinfo{author}{D.~F.~J. Kimball},
  \bibinfo{author}{C.~Pankow}, \bibinfo{author}{M.~P. Ledbetter},
  \bibinfo{author}{P.~Wlodarczyk}, \bibinfo{author}{P.~Wcislo},
  \bibinfo{author}{M.~Pospelov}, \bibinfo{author}{J.~R. Smith},
  \bibinfo{author}{J.~Read}, \bibinfo{author}{W.~Gawlik},
  \bibinfo{author}{D.~Budker},
\newblock \bibinfo{title}{{The Global Network of Optical Magnetometers for
  Exotic physics (GNOME): A novel scheme to search for physics beyond the
  Standard Model}},
\newblock \bibinfo{journal}{Annalen der Physik} \bibinfo{volume}{525}
  (\bibinfo{year}{2013}) \bibinfo{pages}{659--670}.
  \DOIprefix\doi{10.1002/andp.201300061}.
  \href{http://arxiv.org/abs/1303.5524}{{\tt arXiv:1303.5524}}.
\bibitem[{Graham et~al.(2015)Graham, Irastorza, Lamoreaux, Lindner, and van
  Bibber}]{Graham:2015ouw}
\bibinfo{author}{P.~W. Graham}, \bibinfo{author}{I.~G. Irastorza},
  \bibinfo{author}{S.~K. Lamoreaux}, \bibinfo{author}{A.~Lindner},
  \bibinfo{author}{K.~A. van Bibber},
\newblock \bibinfo{title}{{Experimental Searches for the Axion and Axion-Like
  Particles}},
\newblock \bibinfo{journal}{Ann. Rev. Nucl. Part. Sci.} \bibinfo{volume}{65}
  (\bibinfo{year}{2015}) \bibinfo{pages}{485--514}.
  \DOIprefix\doi{10.1146/annurev-nucl-102014-022120}.
  \href{http://arxiv.org/abs/1602.00039}{{\tt arXiv:1602.00039}}.

\end{thebibliography}

\end{document}